\documentclass[12pt]{article}
\usepackage[dvips]{graphicx}
\begin{document}
\title{Randomly evolving trees I}
\author{L. P\'al, \\\footnotesize{KFKI Atomic Energy Research Institute
1525 Budapest, P.O.B. 49. Hungary}}
\date{\footnotesize{May 23, 2002}}

\maketitle
\begin{abstract}
The random process of the tree evolution with continuous time
parameter has been investigated. By introducing the notions of
living and dead nodes a new model of random tree evolution has
been constructed. It is assumed that at $t=0$ the tree is
consisting of a single living node (root) from which the evolution
may begin. This initial state is denoted by ${\mathcal S}_{0}$. At
a certain time instant $\tau \geq 0$ the root produces $\nu \geq
0$ living nodes connected by lines to the root which becomes dead
at the moment of the offspring production. $\tau$ and $\nu$ are
random variables with known distribution functions. In the
evolution process each of the new living nodes independently of
the others evolves further like a root. The state
$\mathcal{S}_{t}$ of the tree at time instant $t > 0$ is defined
by the actual numbers of living and dead nodes. It has been shown
that the generating function of the probability to find a tree in
a given state ${\mathcal S}_{t}$ at $t \geq 0$, assuming that at
$t=0$ it was in the state ${\mathcal S}_{0}$, satisfies a
non-linear integral equation. Analyzing the time dependence of the
expectation value of the number of living nodes it has been
recognized that three completely different types of evolution
exist depending on the average number $q_{1}$ of nodes produced by
one living node. If $q_{1}<1$, then the evolution is subcritical,
if $q_{1} = 1$, then it is critical, while in the case of
$q_{1}>1$ is supercritical. It has been proved that in the case of
subcritical evolution there is a time point at which the variance
of the number of living nodes is maximal. By using the the
generating function of the probability to find at time moment $t
\geq 0$ a given number of nodes in the tree, the time dependence
of the expectation value and the variance of tree size has been
calculated. The joint distribution function of the numbers of
living and dead nodes has been also derived, and the correlation
between these node numbers has been determined as a function of
time parameter. It has been proved that the correlation function
converges to $\sqrt{3}/2$ independently of the distributions of
$\nu$ and $\tau$ when $q_{1} \rightarrow 1$ and $t \rightarrow
\infty$.

\vspace{0.2cm}

\noindent {\bf PACS: 02.50.-r, 02.50.Ey, 05.40.-a}

\end{abstract}

\section{Introduction}

There is a growing interest in randomly evolving graphs since a
large number of practically important problems can be analyzed in
this way. In the last few years so many papers have been published
in this field that any list of references would be far from
complete by any standards. It is very fortunate that recently two
outstanding review papers \cite{barabasi01}, \cite{dorogovtsev01}
have been published and thus the author of the present paper does
not feel obliged to cite the large amount of important references.

There are many different models for the random graphs and it would
be difficult to give a precise classification of
them.\footnote{Throughout this paper we will use the terms "node"
and "line" (sometimes "link") for the terms "vertex" and "edge",
respectively. The degree of a node is defined by the total number
of its links, and the out-degree of a node is nothing but the
number of its outgoing lines.} Probably the oldest model has been
proposed by Erd\H os and R\'enyi \cite{erdos59}, \cite{erdos60},
\cite{erdos61}. In this model the initial state of the graph
consists of fixed number $n$ of nodes that will be randomly
connected during the evolution. The time is discrete and counted
by the number of successively appearing lines. At time $t=1$ one
of the ${n \choose 2}$ possible lines defined by the nodes is
chosen out assuming that each of these lines has the same
probability to be chosen. At time $t=2$ the second line is chosen
out from the ${n \choose 2}-1$ possible lines all these being
equiprobable. Continuing this process at time $t=N$ a random graph
is formed which is consisting of $n$ fixed nodes and $N$ lines
selected randomly. If $n$ is a large positive integer converging
to infinity and $N$ is increasing from $1$ to ${n \choose 2}$ then
the evolution of the graph passes through several (say five)
clearly distinguishable phases.

One has to mention that random networks with fixed number of nodes
usually show the so-called small-world effect, i.e. the average
shortest path is small. This type of graphs can be created by
randomly rewriting some of the lines of a regular ring graph.
Watts and Strogatz \cite{watts98} noticed many interesting
properties of these graphs called small-world networks and a large
number of papers \cite{watts99}, \cite{newman00} was published
recently on this subject.

There is another - more realistic - model \cite{barabasi99} for
the random evolution of graphs. In this model the number of nodes
is not fixed, but to the contrary of the Erd\H os and R\'enyi
model it is increasing in each discrete, equidistant time moment
by addition of a new node connected randomly to one or several of
the nodes already existing in the graph. The probability of a
given connection between the new and one of the old nodes may
depend on the number of links due to the node existing already.
Many variants \cite{krapivsky00}, \cite{dorogovtsev00} of this
scale-free model have been recently investigated by using
different inventive approximations. The one of the most
intensively studied problem is the chracter of the degree
distribution of nodes in graphs evolving randomly with various
preferential linking. The number of exactly solvable (or solved)
problems, however, is not very much, therefore, it seems to be
worthwhile to study models which can be treated exactly.

It is known for long time that the random evolution of a
tree~\footnote{A randomly evolving tree is a connected graph
containing no cycles, and growing from a single node (root)
according to well-defined rules.} corresponds to a Galton-Watson
process \cite{harris63} which describes the formation of a
population which starts at $t=0$ with one individual and in which
at discrete time moments $t=1, 2, \ldots$ each individual
independently of the others produces a random number of offspring
with the same distribution. There are many interesting papers
\cite{lyons02} on this subject but it is hardly to find articles
about the random processes of tree evolution with continuous time
parameter. The purpose of is this paper is the study of these
special random trees with the hope that it brings about new
insight into the nature of branching processes.

This paper is organized as follows. In Section 2 the exact
description of the random process is given, while in Section 3 the
derivations of the basic equations for the generating functions
can be found. The properties of average characteristics
(expectation values, variances and correlation functions) are
discussed in Section 4, and, finally, the conclusions are
summarized in Section 5.

\section{Description of the random process}

We will study the random evolution of such trees which are
consisting of living and dead nodes as well as lines connecting
the nodes. The {\em living node} is unstable, and after a certain
time $\tau$ may become dead. The lifetime $\tau$ is random
variable, and denote by ${\mathcal P}\{\tau \in (t, t+dt)\} =
dT(t)$ the probability of finding $\tau$ in the time interval
$(t,t+dt)$. The {\em dead node} is inactive, cannot change its
state, and is unable to interact with the other nodes.

The evolution process can be described as follows. Let us suppose
that at time instance $t=0$ the tree consists of a single living
node called {\em root} and denote by ${\mathcal S}_{0}$ this state
of the tree. At the end of its life the root creates $\nu =0, 1,
2, \ldots$ new living nodes and after that it becomes immediately
dead. The number of the offspring $\nu$ is random variable, and
the probabilities ${\mathcal P}\{\nu=k\} = f_{k}, \;\; k = 0, 1,
2, \ldots$ are supposed to be known. The new living nodes are
promptly connected to the dead one and independently of the others
each of them can evolve further like a root. It means that the
tree from its {\em initial state} ${\mathcal S}_{0}$ may evolve
further by two mutually exclusive ways:
\begin{itemize}
\item  the root does not die, and thus the tree remains in its
initial state during the whole time interval $(0, t)$, or \item
the root does die in an infinitesimally small subinterval $\;(t',
t' + dt') \in (0, t)\;$ and produces $j \geq 0$ new living nodes.
Each of these nodes will evolve further in the remained time
interval $(t',t)$ independently of the others like the root.
\end{itemize}

\begin{figure}[ht]
\protect \centering{
\includegraphics[height=7.5cm, width=7.5cm] {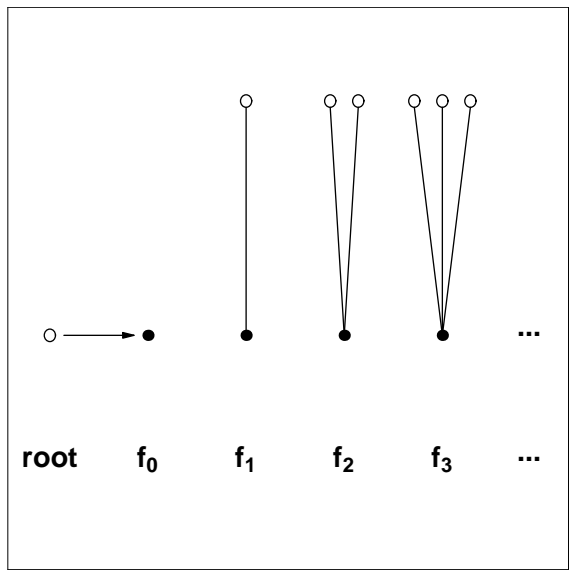}}\protect
\caption{\footnotesize{First step of the tree evolution. The new
state arises from the initial state which consists of a single
root. Four of the possible new states are illustrated in Fig. The
living nodes are denoted by white and the dead ones by black
circles. $f_{0}, f_{1}, f_{2}, f_{3},  \ldots$ are the
probabilities of producing $0, 1, 2, 3, \ldots$ nodes by the dying
root.}} \label{fig1}
\end{figure}

Fig. \ref{fig1} shows the initial state of a tree consisting of a
single root and illustrates four of the possible states which may
be produced by the dying root. The probabilities of producing $0,
1, 2, 3, \ldots$ living nodes are denoted by $f_{0}, f_{1}, f_{2},
f_{3}, \ldots$.

\begin{figure}[ht]
\protect \centering{
\includegraphics[height=8cm, width=8cm] {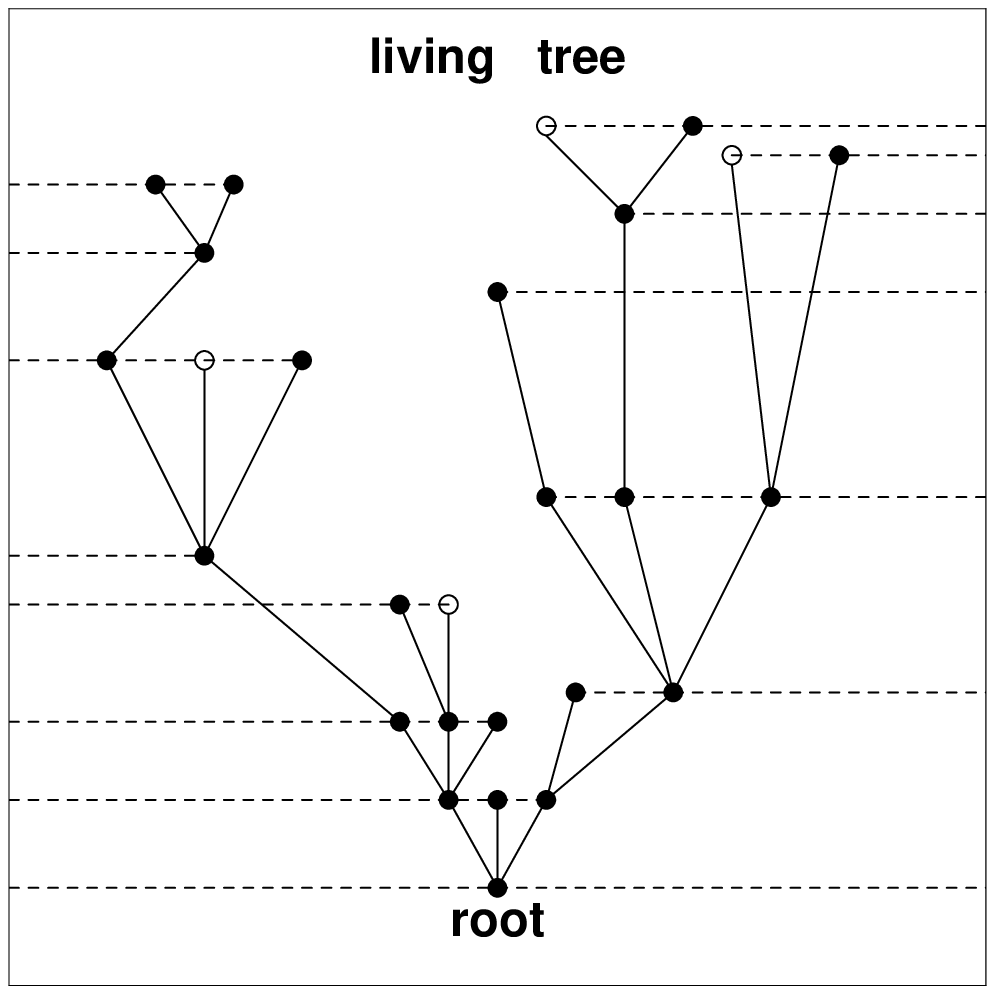}}\protect
\caption{\footnotesize{One of the realizations of a random tree at
the time moment $t > 0$. The nodes denoted by white circle are
capable for the further evolution, while those denoted by black
could not further evolve. The horizontal dashed lines indicate the
time instances at which nodes were produced.}} \label{fig2}
\end{figure}

Fig. \ref{fig2} illustrates one of the realizations of a random
tree at the moment $t > 0)$. One can see in Fig. \ref{fig2} three
white circles denoting the living nodes which are capable for the
further evolution and a large number of the black circles denoting
the dead nodes which are unable to produce new living nodes. It is
to mention that each of the offspring is connected by a line to
the dead precursor. If a dead node has $k$ outgoing lines then it
will be called {\em node of $k$th out-degree}. It is obvious that
a {\em living tree} must have living nodes, while a {\em dead
tree} consists of dead nodes only.

In order to simplify our consideration we will assume that the
distribution function of the lifetime $\tau$ of living nodes is
exponential, i.e.
\begin{equation} \label{1}
dT(t) = e^{-Qt}\;Q\;dt,
\end{equation}
where $1/Q=t_{0}$ is the expectation value of the lifetime. In
this case the evolution becomes a Markov process. It is assumed
that the probabilities
\begin{equation} \label{2}
{\mathcal P}\{\nu = k\} = f_{k}, \;\;\;\;\;\; k = 0, 1, 2, \ldots
\end{equation}
are the same for all living nodes. By introducing the generating
function
\begin{equation} \label{3}
q(z) = \sum_{k=0}^{\infty} f_{k}\;z^{k}
\end{equation}
we can define the factorial moments of $\nu$ in the following way:
\[ q_r = \left[\frac{d^{r}q(z)}{dz^{r}}\right]_{z=1} \;\;\;\;\;\; r
= 1, 2, \ldots\;\;. \] For the expectation value and the variance
of $\nu$ we have
\begin{equation} \label{4}
{\bf E}\{\nu\} = q_1 \;\;\;\;\;\; \mbox{and} \;\;\;\;\;\; {\bf
D}^{2}\{\nu\} = q_2 + q_1 - q_1^{2}.
\end{equation}
For the sake of later use we cite the relation
\begin{equation} \label{5}
{\bf E}\{(\nu-1)^{2}\} = {\bf D}^{2}\{\nu\} + (1-q_1)^{2}.
\end{equation}

In order to characterize the random process of tree evolution we
introduce the {\em random functions} $\mu_{\ell}(t)$ and
$\mu_{d}(t)$ which are the actual numbers of the living and dead
nodes, respectively, at time instant $t \geq 0$. It seems to be
worthwhile to define the random function $\mu(t) = \mu_{\ell}(t) +
\mu_{d}(t)$ which gives the {\em total number of nodes}, i.e. the
{\em actual value of the tree size} at $t \geq 0$. Finally, let us
denote by $\mu_{e}(t)$ the number of lines in a random tree at
time instant $t \geq 0$. It is obvious that the equality
$\mu_{e}(t) = \mu(t) - 1$ is true with probability $1$ for all $t
\geq 0$.

In the next {\em Section} we shall derive the basic equations for
the generating functions of probabilities of the following random
events: $\{\mu_{\ell}(t) = n_{\ell}\}, \;\; \{\mu_{d}(t) =
n_{d}\},\;\; \{\mu_{\ell}(t) = n_{\ell},\; \mu_{d}(t) = n_{d}\}$
and $\{\mu(t) = n\}$, provided that at $t=0$ the tree was in the
state ${\mathcal S}_{0}$.

\section{Derivation of the basic equations}

\subsection{Generating function equations}

\subsubsection{Probability distribution of the number of living nodes}

The living nodes capable to produce new nodes at the end of their
life are responsible for the evolution of the tree. It seems to be
interesting to have an insight into the random process of the
formation of living nodes. Let us denote by
\begin{equation} \label{6}
{\mathcal P}\{\mu_{\ell}(t)=n|{\mathcal S}_{0}\} = p^{(\ell)}(t, n)
\end{equation}
the probability that the number of living nodes $\mu_{\ell}(t)$ is
equal to $n$ at the time instant $t \geq 0$ provided that at $t=0$
the tree was in the state ${\mathcal S}_{0}$. By exploiting the
rules of the random evolution described in {\em Section 2} it can
be easily shown that
\[p^{(\ell)}(t, n) = e^{-Qt}\;\delta_{n,1} + \] \[
Q\;\int_{0}^{t}e^{-Q(t-t')}\left[f_{0}\;\delta_{n,0} +
\sum_{k=1}^{\infty} f_{k}\;\sum_{n_{1}+\cdots+n_{k}=n}\;
\prod_{j=1}^{k}\;p^{(\ell)}(t', n_{j})\right]\;dt', \] and by
using the probability generating function
\[ g^{(\ell)}(t, z) = \sum_{n=0}^{\infty} p^{(\ell)}(t, n)\; z^{n}, \]
we can prove that $g^{(\ell)}(t, z)$ satisfies the integral
equation
\begin{equation} \label{7}
g^{(\ell)}(t, z)= e^{-Qt}\;z + Q\;\int_{0}^{t}e^{-Q(t-t')}\;
q\left[g^{(\ell)}(t', z)\right]\;dt'.
\end{equation}
It is an elementary task to prove that the integral equation
(\ref{7}) is equivalent to the differential equation
\begin{equation}\label{8}
\frac{\partial g^{(\ell)}(t, z)}{\partial t} =
- Q\;g^{(\ell)}(t, z) + Q\;q\left[g^{(\ell)}(t, z)\right],
\end{equation}
with the initial condition
\[ \lim_{t \downarrow 0}g^{(\ell)}(t, z) = z. \]
From the equation (\ref{7}) or (\ref{8}) the time dependence of
the expectation value and the variance of the random function
$\mu_{\ell}(t)$ can be easily calculated.

\subsubsection{Probability distribution of the number of dead nodes}

A living node producing offspring becomes dead, and we are
interested in the probability that the number of dead nodes
$\mu_{d}(t)$ is equal to $n$ at time instant $t \geq 0$ provided
that at $t=0$ the tree was in the state ${\mathcal S}_{0}$. By
introducing the notation
\begin{equation} \label{9}
{\mathcal P}\{\mu_{d}(t)=n|{\mathcal S}_{0}\} = p^{(d)}(t, n),
\end{equation}
we can easily derive the following equation:
\[p^{(d)}(t, n) = e^{-Qt}\;\delta_{n,0} + \] \[
Q\;\int_{0}^{t}e^{-Q(t-t')}\left[f_{0}\;\delta_{n,1} +
\sum_{k=1}^{\infty} f_{k}\;\sum_{n_{1}+\cdots+n_{k}=n-1}\;
\prod_{j=1}^{k}\;p^{(d)}(t', n_{j})\right]\;dt', \] and by using
the probability generating function
\begin{equation} \label{10}
g^{(d)}(t, z) = \sum_{n=0}^{\infty} p^{(d)}(t, n)\; z^{n},
\end{equation}
we can prove that $g^{(d)}(t, z)$ satisfies the integral equation
\begin{equation} \label{11}
g^{(d)}(t, z)= e^{-Qt} + z\;Q\;\int_{0}^{t}e^{-Q(t-t')}\;
q\left[g^{(d)}(t', z)\right]\;dt'.
\end{equation}
It is obvious that the equation (\ref{11}) is equivalent to the
differential equation
\begin{equation}\label{12}
\frac{\partial g^{(d)}(t, z)}{\partial t} =
- Q\;g^{(d)}(t, z) + z\;Q\;q\left[g^{(d)}(t, z)\right],
\end{equation}
with the initial condition
\[ \lim_{t \downarrow 0}g^{(d)}(t, z) = 1. \]

\subsubsection{Probability distribution of the total number of nodes}

Denote by $\mu(t) = \mu_{\ell}(t) + \mu_{d}(t)$ the sum of the
numbers of living and dead nodes at $t \geq 0$. Let us determine
the probability of the random event $\{\mu(t)=n\}$ provided that
at $t=0$ the tree was in the state ${\mathcal S}_{0}$. Introducing
the notation
\begin{equation} \label{13}
{\mathcal P}\{\mu(t)=n|{\mathcal S}_{0}\} = p(t, n)
\end{equation}
and defining the generating function
\begin{equation} \label{14}
g(t, z) = \sum_{n=0}^{\infty} p(t, n)\;z^{n},
\end{equation}
it can be easily shown that $g(t, z)$ satisfies the integral
equation
\begin{equation} \label{15}
g(t, z)= e^{-Qt}\;z + z\;Q\;\int_{0}^{t}e^{-Q(t-t')}\;q\left[g(t',
z)\right]\;dt'.
\end{equation}
For the sake of completeness we are deriving from Eq. (\ref{15})
the equivalent differential equation:
\[ \frac{\partial g(t, z)}{\partial t} = - Q\;g(t, z)
+ Q z\; q\left[g(t, z)\right] \]  the initial condition of which
is
\[ \lim_{t \downarrow 0} g(t, z) = z. \]

\subsubsection{Joint distribution of the numbers of living and dead nodes}

We shall define now the probability
\begin{equation} \label{16}
{\mathcal P}\{\mu_{\ell}(t)=n_{\ell}, \mu_{d}(t)=n_{d}|{\mathcal S}_{0}\} =
p^{(\ell,d)}(t, n_{\ell}, n_{d}),
\end{equation}
where $n_{\ell}$ and $n_{d}$ are a non-negative integers. It is
clear that $p^{(\ell,d)}(t, n_{\ell}, n_{d})$ gives the
probability that at the time instant $t > 0$ the number of living
and that of dead nodes in a randomly evolving tree are equal to
$n_{\ell}$ and $n_{d}$, respectively, provided that at $t=0$ the
tree was in the state ${\mathcal S}_{0}$. By exploiting the rules
of the random evolution described in {\em Section 2} it can be
shown that
\[ p^{(\ell,d)}(t, n_{\ell}, n_{d}) = e^{-Qt}\;\delta_{n_{\ell},1}\;\delta_{n_{d},0} + \]
\[ Q\;\int_{0}^{t} e^{-Q(t-t')}\left\{f_{0}\;\delta_{n_{\ell},0}\;\delta_{n_{d},1} +
\sum_{k=1}^{\infty}f_{k}\;V^{(\ell,d)}(t', n_{\ell}, n_{d})\right\}\;dt, \]
where
\[ V^{(\ell,d)}(t', n_{\ell}, n_{d}) =
\sum_{n_{\ell,1}+\cdots+n_{\ell,k}=n_{\ell}}\;\sum_{n_{d,1}+\cdots+n_{d,k}=n_{d}-1}
\prod_{j=1}^{k}p^{(\ell,d)}(t', n_{\ell,j}, n_{d,j}). \]
Introducing the probability generating function
\begin{equation} \label{17}
g^{(\ell,d)}(t, z_{\ell}, z_{d}) = \sum_{n_{\ell}=0}^{\infty}\;
\sum_{n_{d}=0}^{\infty} p^{(\ell,d)}(t, n_{\ell},
n_{d})\;z_{\ell}^{n_{\ell}}\;z_{d}^{n_{d}},
\end{equation}
and taking into account the generating function defined by Eq.
(\ref{3}) we obtain the {\em fundamental equation} in the form:
\begin{equation} \label{18}
g^{(\ell,d)}(t, z_{\ell}, z_{d})= e^{-Qt}\;z_{\ell} + z_{d}\;Q\;\int_{0}^{t}
e^{-Q(t-t')}\;q\left[g^{(\ell,d)}(t', z_{\ell}, z_{d})\right]\;dt',
\end{equation}
which can be rewritten as a nonlinear differential equation
\begin{equation} \label{19}
\frac{\partial g^{(\ell,d)}(t, z_{\ell}, z_{d})}{\partial t} =
- Q\;g^{(\ell,d)}(t, z_{\ell}, z_{d}) +
z_{d}\;Q\;q\left[g^{(\ell,d)}(t, z_{\ell}, z_{d})\right]
\end{equation}
with initial condition
\[ \lim _{t \downarrow 0} g^{(\ell,d)}(t, z_{\ell}, z_{d}) = z_{\ell}. \]
One has to note that the former equations (\ref{7}) and (\ref{11})
can be immediately derived from the integral equation (\ref{18}).

\subsubsection{Joint distribution of the numbers of living nodes and lines}

As it has been already mentioned the living nodes are responsible
for the evolution of the tree. It seems to be interesting to have
a deeper insight into the interplay between the living nodes and
lines during the tree evolution. For this purpose let us determine
the probability that the number of living nodes $\mu_{\ell}(t)$
born in the time interval $(0, t)$ and the number of lines
$\mu_{e}(t)$ produced in the same time interval are equal to
$n_{\ell}$ and  $n_{e}$, respectively, provided that at the time
instant $t=0$ the tree was in the state ${\mathcal S}_{0}$.
Introducing the notation
\begin{equation} \label{20}
{\mathcal P}\{\mu_{\ell}(t)=n_{\ell}, \mu_{e}(t)=n_{e}|{\mathcal
S}_{0}\} = p^{(\ell,e)}(t, n_{\ell}, n_{e}),
\end{equation}
we can easily derive the backward equation
\[p^{(\ell,e)}(t, n_{\ell}, n_{e}) = e^{-Qt}\;\delta_{n_{\ell},1}\;\delta_{n_{e},0} + \]
\[ Q\;\int_{0}^{t}e^{-Q(t-t')}\left[f_{0}\;\delta_{n_{\ell},0}\;\delta_{n_{e},0} +
\sum_{k=1}^{\infty} f_{k}\;M_{k}(t',n_{\ell}, n_{e})\right]\;dt',
\] where
\[ M_{k}(t',n_{\ell}, n_{e}) = \sum_{n_{\ell,1}+\cdots+n_{\ell,k}=n_{\ell}}\;
\sum_{n_{e,1}+\cdots+n_{e,k}=n_{e}-k}\;\prod_{j=1}^{k}\;p^{(\ell,e)}(t',
n_{\ell,j}, n_{e,j}).\] The probability generating function
\begin{equation} \label{21}
g^{(\ell,e)}(t, z_{\ell}, z_{e}) =
\sum_{n_{\ell}=0}^{\infty}\;\sum_{n_{e}=0}^{\infty}
p^{(\ell,e)}(t, n_{\ell},
n_{e})\;z_{\ell}^{n_{\ell}}\;z_{e}^{n_{e}},
\end{equation}
we obtain
\begin{equation} \label{22}
g^{(\ell,e)}(t, z_{\ell}, z_{e}) = e^{-Qt}\;z_{\ell} +
Q\;\int_{0}^{t}e^{-Q(t-t')}\;q[z_{e}\;g^{(\ell,e)}(t', z_{\ell},
z_{e})]\;dt'.
\end{equation}
This is an important equation from which one can easily derive
relations for the time dependence of factorial moments
$\mu_{\ell}(t)$, and $\mu_{e}(t)$, and calculate the correlation
function between them. The particular variant of the Eq.
(\ref{22}), namely:
\[ g^{(\ell,e)}(t, z_{\ell}=z, z_{e}=1) = g^{(\ell)}(t, z) = e^{-Qt}\;z +
Q\;\int_{0}^{t}e^{-Q(t-t')}\;q\left[g^{(\ell)}(t', z)\right]\;dt'
\]
is equivalent to Eq. (\ref{7}), and by substitutions $z_{\ell}=1$
and $z_{e}=z$ one can obtain the integral equation
\[ g^{(\ell,e)}(t, z_{\ell}=1, z_{e}=z) = g^{(e)}(t, z) = e^{-Qt} +
Q\;\int_{0}^{t}e^{-Q(t-t')}\;q\left[z\;g^{(e)}(t', z)\right] \;dt'
\]
determining the generating function $g^{(e)}(t, z) = {\bf
E}\{z^{\mu_{e}}\}$.

The differential equation corresponding to (\ref{22}) has the
form:
\[ \frac{\partial g^{(\ell,e)}(t, z_{\ell}, z_{e})}{\partial t} =
-Q\;g^{(\ell,e)}(t, z_{\ell}, z_{e}) +
Q\;q\left[z_{e}\;g^{(\ell,e)}(t, z_{\ell}, z_{e})\right], \] with
the initial condition
\[ \lim_{t \downarrow 0}g^{(\ell,e)}(t, z_{\ell}, z_{e}) =
z_{\ell}. \]

\subsubsection{Remarks on generating function
equations}

In this {\em Section} we derived integral equations for the
following generating functions:
\[ g^{(\ell)}(t, z) = {\bf E}\left\{z^{\mu_{\ell}(t)}\right\}, \;\;\;\;\;\;
g^{(d)}(t, z) = {\bf E}\left\{z^{\mu_{d}(t)}\right\}, \;\;\;\;\;\;
g(t, z) = {\bf E}\left\{z^{\mu(t)}\right\} \] and
\[ g^{(\ell, d)}(t, z_{\ell}, z_{d}) = {\bf E}\left\{z_{\ell}^{\mu_{\ell}(t)}
\;z_{d}^{\mu_{d}(t)}\right\}, \;\;\;\; g^{(\ell, e)}(t, z_{\ell},
z_{e}) = {\bf E}\left\{z_{\ell}^{\mu_{\ell}(t)}
\;z_{e}^{\mu_{e}(t)}\right\}. \] For the sake of simplicity let us
denote by $\gamma(t, z)$~\footnote{The variable $z$ may have two
components.} any of these generating functions. It can be shown
\cite{sevast'yanov71} that the integral equation for $\gamma(t,
z)$ has just one solution satisfying the conditions
\[ |\gamma(t, z)| \leq 1, \;\;\;\; \forall \;\;|z|\leq 1 \;\;\;\;
\mbox{and} \;\;\;\; \lim_{z \uparrow 1}\;\; \gamma(t, z) = 1. \]
It is important to mention that if $\lim_{z \uparrow 1}\;
\gamma(t, z) < 1$, then the number of the tree elements (nodes,
lines) increases so rapidly that it reach $\infty$ in a finite
time interval. In our further considerations, however, this {\em
explosion-like evolution} will be excluded.

\section{Average characteristics}

\subsection{Expectation values and variances}

One of the simplest way to study the characteristic properties of
tree evolution is to analyze the {\em time dependence of factorial
moments} of the following random functions:
\[ \mu_{\ell}(t), \;\;\;\; \mu_{d}(t), \;\;\;\; \mbox{and} \;\;\;\;\mu(t). \]
For the sake of simplicity let us denote by  $\chi(t)$ any of
these random functions and define the generating function
\begin{equation} \label{23}
\gamma(t, z) = {\bf E}\{z^{\chi(t)}\}.
\end{equation}
According to the Abel's theorem the $j$th factorial moment
\begin{equation} \label{24}
{\bf E}\{\chi(t)\;[\chi(t)-1]\;\cdots \;[\chi(t)-j+1]\} = m_{j}(t)
= \left[\frac{\partial^{j} \gamma(t, z)}{\partial
z^{j}}\right]_{z=1},
\end{equation}
exists if and only if the series
\[ \sum_{n=0}^{\infty} {\mathcal P}\{\chi(t)=n|{\mathcal S}_{0}\}\;n^{j},
\;\;\;\; j=1, 2, \ldots , \] converges uniformly at any finite
time instant $t > 0$. It is well-known  that the simple moments
\[ {\bf E}\{[\chi(t)]^{j}\} = m^{(j)}(t) \]
can be expressed by the factorial moments \cite{lpal95} by using
the relation
\begin{equation} \label{25}
m^{(j)}(t) = \sum_{i=0}^{j} {\mathcal S}(j, i)\;m_{i}(t), \;\;\;\;
j=1, 2, \ldots \;,
\end{equation}
where ${\mathcal S}(j, i), \;\; i=0, 1, \ldots, j$ are the
Stirling numbers of the second kind.

In the further considerations we are going to deal with two
important average characteristics of the random function
$\chi(t)$, namely with the expectation value ${\bf E}\{\chi(t)\} =
m_{1}(t)$ and the variance ${\bf D}^{2}\{\chi(t)\} = m_{2}(t) +
m_{1}(t)\;[1 - m_{1}(t)]$.

In order to calculate these average characteristics we do not have
to know the distribution function of $\nu$, it is enough to know
the values of the first and second factorial moments of $\nu$
only. In this case the distribution of $\nu$ we call {\em
arbitrary} ({\bf a}). However, if we are interested in solving the
generating function equations, then we should have the exact form
of the distribution of $\nu$. In the sequel, we will use two
simple distributions, namely the Poisson ({\bf p})
\[{\mathcal P}\{\nu=k\} = e^{-q1}\;\frac{q_{1}^{k}}{k!}, \]
and the geometric ({\bf g})
\[{\mathcal P}\{\nu=k\} = \frac{1}{1+q_{1}}\;\left(\frac{q_{1}}{1+q_{1}}\right)^{k}. \]
distributions. In the case of the Poisson distribution we have
\[ {\bf E}\{\nu\} = q_{1} \;\;\;\;\;\; \mbox{and} \;\;\;\;\;\;
{\bf D}^{2}\{\nu\} = q_{1}, \] while in that of the geometric one
\[ {\bf E}\{\nu\} = q_{1} \;\;\;\;\;\; \mbox{and} \;\;\;\;\;\; {\bf D}^{2}\{\nu\} =
q_{1}(1+q_{1}). \]

\subsubsection{Living nodes}

Let as investigate the expectation value and the variance of the
number of living nodes at the time instant $t > 0$. The equation
for the expectation value ${\bf E}\{\mu_{\ell}(t)\} =
m_{1}^{(\ell)}(t)$ can be immediately derived from the Eq.
(\ref{8}). We obtain
\[ m_{1}^{(\ell)}(t) = e^{-Qt} + Q\;q_{1}\;\int_{0}^{t} e^{-Q(t-t')}
\;m_{1}^{(\ell)}(t')\;dt', \] which is a typical {\em renewal
equation}. After elementary calculations the solution is obtained
in the form:
\begin{equation} \label{26}
m_{1}^{(\ell)}(t) = e^{- (1-q_{1})Qt},
\end{equation}
from which we see that the average number of the living nodes has
the following asymptotic properties:
\begin{equation} \label{27}
\lim_{t \rightarrow \infty} = \left\{ \begin{array}{ll}
           0, & \mbox{if $q_1<1$,} \\
           1, & \mbox{if $q_1=1$,} \\
      \infty, & \mbox{if $q_1>1$. }
        \end{array} \right.
\end{equation}
The evolution of the tree will be called {\em subcritical}, if
$q_1 < 1$, {\em critical}, if $q_1 = 1$, and {\em supercritical},
if $q_1 > 1$.

In order to calculate the {\em variance} ${\bf
D}^{2}\{\mu_{\ell}(t)\}$ we need the second factorial moment
$m_{2}^{(\ell)}(t)$ which can obtained by solving the integral
equation
\[ m_{2}^{(\ell)}(t) = Q\;\int_{0}^{t} e^{-Q(t-t')}
\left\{q_1\;m_{2}^{(\ell)}(t') +
q_2\;\left[m_{1}^{(\ell)}(t')\right]^{2}\right\}\;dt'. \]
It is easy to show that
\begin{equation} \label{28}
m_{2}^{(\ell)}(t) = \left\{ \begin{array}{ll}
        \frac{q_2}{1 - q_1}\;[1 - e^{-(1-q_1)Qt}]\;e^{-(1-q_1)t}, &
        \mbox{if $q_1 \not= 1$,} \\
        \mbox{ } & \mbox{ } \\
        q_2\;Qt, & \mbox{if $q_1=1$,}
                   \end{array} \right.
\end{equation}
and by using the relation
\[ {\bf D}^{2}\{\mu_{\ell}(t)\} = m_{2}^{(\ell)}(t) + m_{1}^{(\ell)}(t)\;[1 -
m_{1}^{(\ell)}(t)] \] we have
\begin{equation} \label{29}
{\bf D}^{2}\{\mu_{\ell}(t)\} = \left\{ \begin{array}{ll}
        \frac{{\bf E}\{(\nu-1)^{2}\}}{1 - q_{1}}\;[1 - e^{-(1-q_1)Qt}]
        \; e^{-(1-q_1)Qt}, & \mbox{if $q_1 \not= 1$,} \\
        \mbox{ } & \mbox{ } \\
        q_2\;Qt, & \mbox{if $q_1=1$,}
                   \end{array} \right.
\end{equation}
where ${\bf E}\{(\nu-1)^{2}\} = (1 - q_{1})^{2} + {\bf
D}^{2}\{\nu\}$. In the case of subcritical ($q_1<1$) evolution the
variance has a maximum at
\[ x_{max} = Q\;t_{max} = \frac{\log 2}{1-q_1},  \]
which is independent of the form of the distribution of $\nu$.
\begin{figure}[ht]
\protect \centering{
\includegraphics[height=7.5cm, width=12cm] {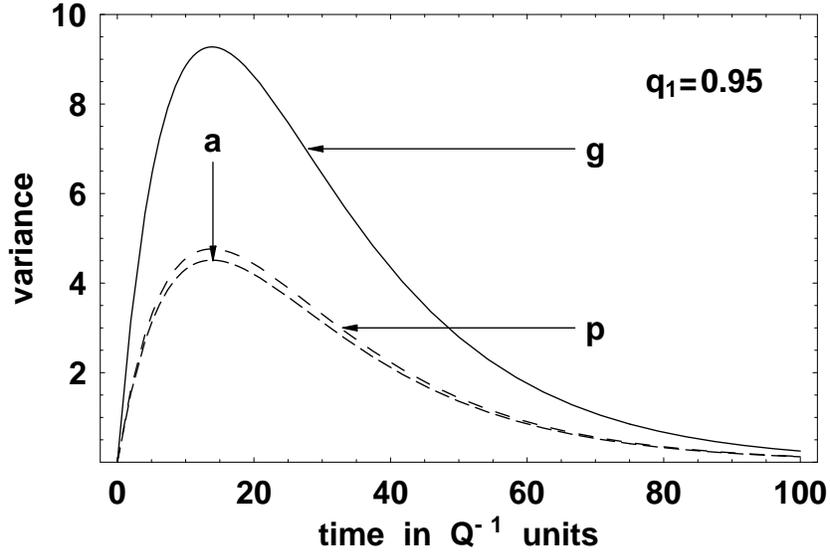}}\protect
\caption{\footnotesize{The variance of the number of living nodes
vs. time parameter $Qt$ in subcritical ($q_{1}=0.95$) evolution.
The curves {\bf g, p}, and {\bf a} correspond to the geometric,
Poisson, and the arbitrary distributions of $\nu$, respectively.
In the case of the arbitrary distribution the value ${\bf
D}^{2}\{\nu\}=0.9$ has been chosen.}} \label{fig3}
\end{figure}
The variance vs. $Qt$ is shown in Fig. \ref{fig3}. The curves {\bf
g, p}, and {\bf a} correspond to the geometric, Poisson, and
arbitrary distributions of $\nu$, respectively. In the case of the
arbitrary distribution the value ${\bf D}^{2}\{\nu\}=0.9$ has been
chosen. The sites of maxima of curves ${\bf g, p}$ and ${\bf a}$
are the same, but the heights of the maxima depend, however, on
the character of the distribution of $\nu$.

The {\em relative variance} of the number of living nodes at the
time instant $t$ can be written in the form:
\begin{equation} \label{30}
\frac{{\bf D}^{2}\{\mu_{\ell}(t)\}}{\left[{\bf E}\{\mu_{\ell}(t)\}\right]^{2}} =
\left\{ \begin{array}{ll}
        \frac{{\bf E}\{(\nu-1)^{2}\}}{1 - q_{1}}\;[e^{(1-q_1)Qt} - 1], &  \mbox{if $q_1                                                 \not= 1$,} \\
        \mbox{ } & \mbox{ } \\
        q_2\;Qt, & \mbox{if $q_1=1$,}
                   \end{array} \right.
\end{equation}
from which it is evident that the relative variance is increasing
with $t$ to $\infty$ in the subcritical and critical evolutions
($q_1 \leq 1$), and converging to the saturation value
\[ \frac{{\bf E}\{(\nu-1)^{2}\}}{q_{1} - 1},  \]
if the evolution is supercritical ($q_{1}>1$). It is trivial but
interesting to note that the standardized expectation value of
$\mu_{\ell}(t)$ has the form:
\begin{equation} \label{31}
\frac{{\bf E}\{\mu_{\ell}(t)\}}{{\bf D}\{\mu_{\ell}(t)\}} = \left\{\begin{array}{ll}
        \left(\frac{1-q_{1}}{{\bf E}\{(\nu-1)^{2}\}}\right)^{1/2}
        \;\left[e^{(1-q_{1})Qt} - 1\right]^{-1/2}, & \mbox{if $q_{1} \neq 1$,} \\
        \mbox{ } & \mbox{ } \\
        \frac{1}{\sqrt{q_{2}}}\;(Qt)^{-1/2}, & \mbox{if $q_{1}=1$.}
                                  \end{array} \right.
\end{equation}
If the evolution is critical, then the standardized expectation
value of the number of living nodes converges to zero as
$(Qt)^{-1/2}$ when $t \Rightarrow \infty$.

\subsubsection{Dead nodes}

It is obvious that the number of dead nodes $\mu_{d}(t)$ in an
evolving tree is a non-decreasing random function of time. From
Eq. (\ref{11}) we obtain for the expectation value ${\bf
E}\{\mu_{d}(t)\} = m_{1}^{(d)}(t)$ the following equation:
\begin{equation} \label{32}
m_{1}^{(d)}(t) = 1 - e^{-Qt} + q_{1}\;Q\;\int_{0}^{t} e^{-Q(t-t')}\;m_{1}^{(d)}(t')\;dt',
\end{equation}
the solution of that can be written in the form:
\begin{equation} \label{33}
m_{1}^{(d)}(t) = \left\{ \begin{array}{ll}
\frac{1}{1-q_{1}}\;\left[1 - e^{-(1-q_{1})Qt}\right], & \mbox{if $q_{1} \neq 1$,} \\
\mbox{ } & \mbox{ } \\
Qt, &  \mbox{if $q_{1} = 1$.}
\end{array} \right.
\end{equation}
We see that the expectation value of dead nodes converges to $(1 -
q_{1})^{-1}$ in subcritical and becomes infinite in critical and
supercritical evolution when $t \Rightarrow \infty$. It is
remarkable that the increase of $m_{1}^{(d)}(t)$ with time $t$ is
linear in critical state.

In order to calculate the variance ${\bf D}^{2}\{\mu_{d}(t)\}$ of
the number of dead nodes at the time instant $t \geq 0$ we have to
solve the integral equation
\[ m_{2}^{(d)}(t) = q_{1}\;Q\;\int_{0}^{t} e^{-Q(t-t')}\;m_{2}^{(d)}(t')\;dt' +  \]
\begin{equation} \label{34}
Q\;\int_{0}^{t} e^{-Q(t-t')}\;\left\{2 q_{1}\;m_{1}^{(d)}(t') +
q_{2}\;\left[m_{1}^{(d)}(t')\right]^{2}\right\}\;dt',
\end{equation}
which can be derived from Eq. (\ref{11}) by using the formula
\begin{equation} \label{35}
m_{2}^{(d)}(t) = \left[\frac{\partial^{2} g^{(d)}(t, z)}{\partial z^{2}}\right]_{z=1}.
\end{equation}
After elementary calculations we obtain the solution in the form:
\[ m_{2}^{(d)}(t) =a_{1}^{(d)}\;[1 - e^{-(1-q_{1})Qt}] + \]
\begin{equation} \label{36}
a_{2}^{(d)}\;Qt\;e^{-(1-q_{1})Qt} +
a_{3}^{(d)}\;e^{-(1-q_{1})Qt}\; [1 - e^{-(1-q_{1})Qt}],\;\;\;
\mbox{if} \;\;\; q_{1} \neq 1,
\end{equation}
where
\[ a_{1}^{(d)} = \frac{1}{(1-q_{1})^{2}}\left(2\;q_{1} +
\frac{q_{2}}{1-q_{1}}\right), \;\;\;\; a_{2}^{(d)} =
-\frac{2}{1-q_{1}}\left(q_{1} + \frac{q_{2}}{1-q_{1}}\right), \]
and \[  a_{3}^{(d)} = \frac{q_{2}}{(1-q_{1})^{3}}. \] If $q_{1} =
1$, then we have
\begin{equation} \label{37}
m_{2}^{(d)}(t)  = (Qt)^{2} + \frac{1}{3}\;q_{2}\;(Qt)^{3}.
\end{equation}
By using the expression
\[ {\bf D}^{2}\{\mu_{d}(t)\} = m_{2}^{(d)}(t) +  m_{1}^{(d)}(t)\;[1 -
m_{1}^{(d)}(t)] \] if $q_{1} \neq 1$, then we find
\[ {\bf D}^{2}\{\mu_{d}(t)\} = v_{1}^{(d)}\;[1 - e^{-(1-q_{1})Qt}] +  \]
\begin{equation} \label{38}
v_{2}^{(d)}\;Qt \;e^{-(1-q_{1})Qt} +
v_{3}^{(d)}\;e^{-(1-q_{1})Qt}\; [1 - e^{-(1-q_{1})Qt}],
\end{equation}
where
\[ v_{1}^{(d)} =  \frac{{\bf D}^{2}\{\nu\}}{(1-q_{1})^{3}},
\;\;\;\; v_{2}^{(d)} = - 2\;\frac{{\bf
D}^{2}\{\nu\}}{(1-q_{1})^{2}},  \;\;\;\; v_{3}^{(d)} = \frac{{\bf
E}\{(\nu-1)^{2}\}}{(1-q_{1})^{3}}, \] and if $q_{1} = 1$, i.e. the
evolution is critical, then
\begin{equation} \label{39}
{\bf D}^{2}\{\mu_{d}(t)\}  =  Qt + \frac{1}{3}\;q_{2}\;(Qt)^{3}.
\end{equation}
One can see that the limit values of the variance of the number of
dead nodes are given by the formula
\[ \lim_{t \rightarrow \infty}{\bf D}^{2}\{\mu_{d}(t)\} =  \left\{ \begin{array}{ll}
\frac{{\bf D}^{2}\{\nu\}}{(1-q_{1})^{3}}, & \mbox{if $q_{1}<1$,} \\
            \mbox{} & \mbox{} \\
              \infty, &  \mbox{if $q_{1} \geq 1$.}
             \end{array} \right. \]

\subsubsection{Total number of nodes}

For many purposes, it may be important to know how the expectation
value and the variance of the random function
\begin{equation} \label{40}
\mu(t) = \mu_{\ell}(t) + \mu_{d}(t)
\end{equation}
depend on the time. By using Eq. (\ref{15}) one can derive
integral equations the factorial moments
\begin{eqnarray}
m_{1}(t) = {\bf E}\{\mu(t)\} & = & \left[\frac{\partial
g(t, z)}{\partial z}\right]_{z=1}, \label{41} \\
m_{2}(t) = {\bf E}\{\mu(t)[\mu(t)-1]\} & = &
\left[\frac{\partial^{2} g(t, z)}{\partial z^{2}}\right]_{z=1}.
\label{42}
\end{eqnarray}
After a brief calculation we obtain that
\begin{equation} \label{43}
m_{1}(t) = 1 + q_{1}\;Q\;\int_{0}^{t}
e^{-Q(t-t')}\;m_{1}(t')\;dt',
\end{equation}
and
\[ m_{2}(t) = q_{1}\;Q\;\int_{0}^{t} e^{-Q(t-t')}\;m_{2}(t')\;dt' + \]
\begin{equation} \label{44}
Q\;\int_{0}^{t} e^{-Q(t-t')}\;\left\{2q_{1}\;m_{1}(t') +
q_{2}\;\left[m_{1}(t')\right]^{2}\right\}\;dt'.
\end{equation}
The solution of Eq. (\ref{43}) has the form:
\begin{equation} \label{45}
m_{1}(t) = \left\{ \begin{array}{ll}
   (1-q_{1})^{-1}\;\left[1 - q_{1}\;e^{-(1-q_{1}) Qt}\right], & \mbox{if $q_{1} \neq 1$,} \\
      \mbox{ } & \mbox{ } \\
   1 +Qt & \mbox{if $q_{1} = 1$,} \end{array} \right.
\end{equation}
and looking at Eqs. (\ref{26}) and (\ref{33}) we can see the
trivial identity
\[ m_{1}(t) = m_{1}^{(\ell)}(t) + m_{1}^{(d)}(t) \]
to be valid. For the purpose of the calculation of the variance
\[ {\bf D}^{2}\{\mu(t)\} = m_{2}(t) + m_{1}(t)[1 -
m_{1}(t)]  \]  we have to obtain the solution of Eq. (\ref{44}).
By having this solution we can immediately write:
\[ {\bf D}^{2}\{\mu(t)\} = \] \[ \frac{{\bf D}^{2}\{\nu\}}{(1-q_{1})^{3}}
\;\left[1 - e^{-(1-q_{1}) Qt}\right] -2q_{1}\;\frac{{\bf
D}^{2}\{\nu\}}{(1-q_{1})^{2}}\;Qt\;e^{-(1-q_{1}) Qt} + \]
\begin{equation} \label{46}
q_{1}^{2}\; \frac{{\bf
E}\{(\nu-1)^{2}\}}{(1-q_{1})^{3}}\;e^{-(1-q_{1}) Qt}\; \left[1 -
e^{-(1-q_{1}) Qt}\right], \;\;\; \mbox{if} \;\;\; q_{1} \neq 1,
\end{equation}
and in the case of critical evolution, i.e. if $q_{1}=1$, than we
have a simple formula:
\begin{equation} \label{47}
{\bf D}^{2}\{\mu(t)\} = (1+q_{2})\;Qt + q_{2}\;(Qt)^{2} +
\frac{1}{3}\;q_{2}\;(Qt)^{3}.
\end{equation}

The {\em average tree size} at a given time instant $t>0$ can be
characterized by the expectation value of the total number of
nodes $\mu(t)$, but it is to mention that the actual sizes of
randomly evolving trees may fluctuate significantly around the
expectation value ${\bf E}\{\mu(t)\}$. This fluctuation can be
measured by the {\em relative variance}
\[ \frac{{\bf D}^{2}\{\mu(t)\}}{\left[{\bf E}\{\mu(t)\}\right]^{2}} =
 {\bf VR}\{\mu(t)\} \]
the time dependence of which shows interesting features. We see
from Eq. (\ref{45}) \hspace{0.2cm} that the average tree size
converges to infinity with $\;t \Rightarrow \infty\;$, if $q_{1}
\geq 1$, and to $(1-q_{1})^{-1}$, if $q_{1}<1$.
\begin{figure}[!t]
\protect \centering{
\includegraphics[height=7cm, width=10cm]{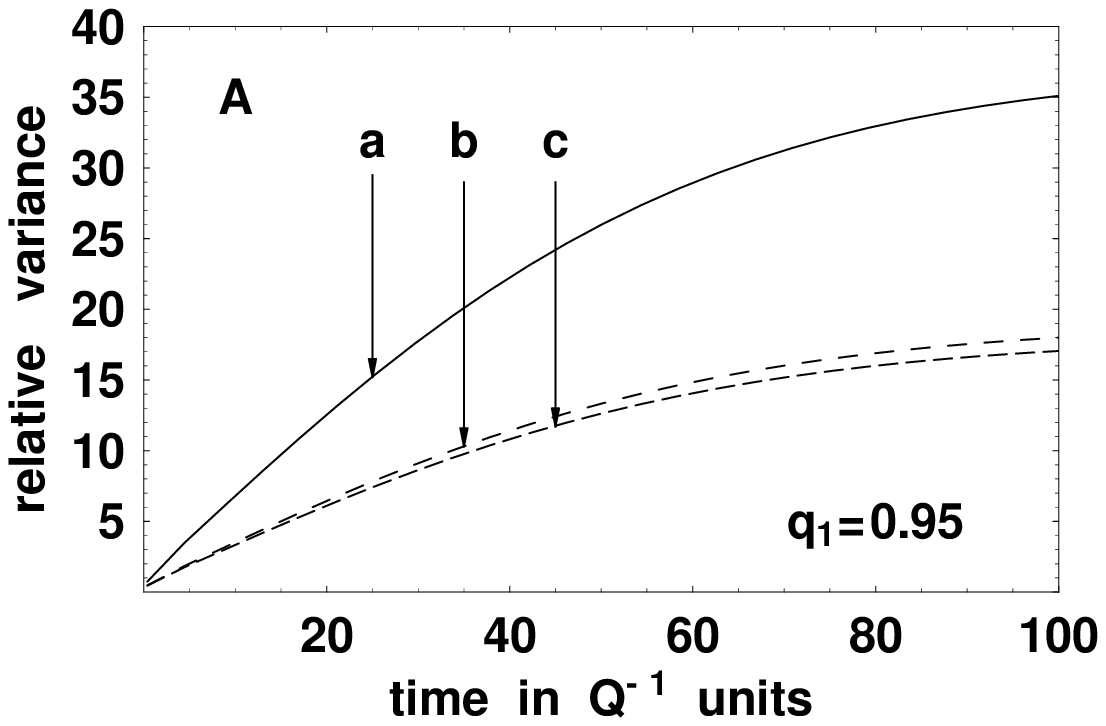}}\protect
\vskip 0.2cm \protect \centering{
\includegraphics[height=7cm, width=10cm] {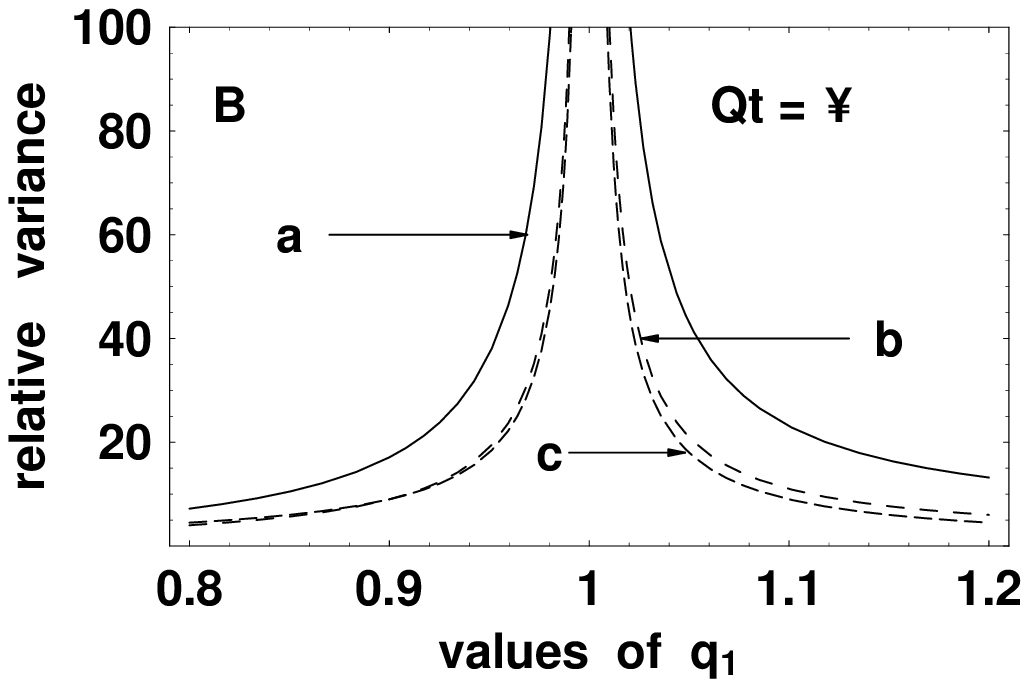}} \protect
\caption{\footnotesize{Fig. {\bf A} shows the relative variance of
the total number of nodes vs. time parameter $Qt$. In Fig. {\bf B}
it can be seen at $Qt=\infty$ the dependence of relative variances
on parameter $q_{1}$. In both Figs. the curves {\bf g, p}, and
{\bf a} correspond to the geometric, Poisson, and the arbitrary
distributions of $\nu$, respectively. In the case of the arbitrary
distribution ${\bf D}^{2}\{\nu\}=0.9$ has been chosen for the
calculations.}} \label{fig4}
\end{figure}
On the contrary, the relative variance remains finite when $t
\Rightarrow \infty$ in the case of $q_{1} \neq 1$, and if and only
if the evolution is critical, i.e., if $q_{1}=1$, then can be
observed the relative variance to become infinite when $t
\Rightarrow \infty$.

In the part {\bf A} of Fig. \ref{fig4} three curves of the
relative variance of $\mu(t)$ versus $Qt$ are plotted when
$q_{1}=0.95$. In the part {\bf B} the dependence of the relative
variance at $Qt=\infty$ can be seen on the parameter $q_{1}$. In
both parts of Fig. the curves {\bf g, p}, and {\bf a} correspond
to the geometric, Poisson, and the arbitrary distributions of
$\nu$, respectively. At $q_{1}=1$ one can observe {\em singularity
in the relative variance}, and thus one has to conclude that the
average tree size at large $Qt$ cannot characterize the random
evolution in critical state.

\subsection{Covariances and correlations}

The study of how the random functions $\mu_{\ell}(t), \mu_{d}(t)$
and $\mu_{e}(t)$ are related involves the analysis of at least two
correlation functions. In the following we will investigate
correlations between the living and dead nodes, as well as the
lines and living nodes.

\subsubsection{Living and dead nodes}

In {\em Sub-subsection 3.1.4} we derived an integral equation [see
Eq. (\ref{18})] for the generating function $g^{(\ell,d)}(t,
z_{\ell}, z_{d})$ of the joint distribution $p^{(\ell, d)}(t,
n_{\ell}, n_{d})$ of random functions $\mu_{\ell}(t)$ and
$\mu_{d}(t)$. It is clear that $p^{(\ell, d)}(t, n_{\ell}, n_{d})$
gives the probability that at time instant $t \geq 0$ the number
of living and that of dead nodes in a randomly evolving tree are
equal to $n_{\ell}$ and $n_{d}$, respectively, provided that at
$t=0$ the tree was in the state ${\mathcal S}_{0}$.

{\bf Covariance function.} \hspace{0.2cm} In order to calculate
the {\em covariance function}
\begin{equation} \label{48}
{\bf Cov}\{\mu_{\ell}(t) \mu_{d}(t)\} = {\bf E}\{\mu_{\ell}(t)
\mu_{d}(t)\} - {\bf E}\{\mu_{\ell}(t)\}{\bf E}\{\mu_{d}(t)\}
\end{equation}
we need the equation determining the moment
\begin{equation} \label{49}
{\bf E}\{\mu_{\ell}(t) \mu_{d}(t)\} = \left[\frac{\partial^{2}
g^{(\ell,d)}(t, z_{\ell}, z_{d})}{\partial z_{\ell} \partial
z_{d}}\right]_{z_{\ell}=z_{d}=1} = m_{1,1}^{(\ell,d)}(t).
\end{equation}
The equation can be written in the form:
\[ m_{1,1}^{(\ell,d)}(t) = Q\;q_{1}\;\int_{0}^{t}
e^{-Q(t-t')}\;m_{1,1}^{(\ell,d)}(t')\;dt' + \]
\begin{equation} \label{50}
Q\;\int_{0}^{t} e^{-Q(t-t')}\left[q_{1}\;m_{1}^{(\ell)}(t') +
q_{2}\;m_{1}^{(\ell)}(t')\;m_{1}^{(d)}(t')\right]\;dt'.
\end{equation}
By using the Eqs. (\ref{26}) and (\ref{33}) for
$m_{1}^{(\ell)}(t)$ and $m_{1}^{(d)}(t)$, respectively, and
introducing the notation $\alpha = (1-q_{1})\;Q$, it can be proved
that
\[ m_{1,1}^{(\ell,d)}(t) = \frac{{\bf
D}^{2}\{\nu\}}{1-q_{1}}\;Qt\;e^{-\alpha t} -
q_{2}\;\frac{1}{(1-q_{1})^{2}}\;e^{- \alpha t}\;(1 - e^{- \alpha
t}), \] is the unique solution of the integral equation (\ref{50})
$\forall \;\; q_{1} \neq 1$. After a brief calculation we have
\begin{equation} \label{51}
{\bf Cov}\{\mu_{\ell}(t) \mu_{d}(t)\}  = \frac{{\bf
D}^{2}\{\nu\}}{1-q_{1}}\;Qt\;e^{-\alpha t} - \left[1 + \frac{{\bf
D}^{2}\{\nu\}}{(1-q_{1})^{2}}\right]\;e^{- \alpha t}\;(1 - e^{-
\alpha t}),
\end{equation}
\[ \mbox{if} \;\;\;\;\;\; q_{1} \neq 1. \]
It is easy to show that in the case of critical evolution, i.e.,
if $q_{1}=1$, then
\begin{equation} \label{52}
{\bf Cov}\{\mu_{\ell}(t) \mu_{d}(t)\} = \frac{1}{2}\;{\bf
D}^{2}\{\nu\}\;(Qt)^{2}.
\end{equation}

{\bf Remark} Now, we would like to show that this result can be
derived without solving the integral equation (\ref{50}). Since
$\mu(t) = \mu_{\ell}(t) + \mu_{d}(t)$ we can write that
\[ {\bf D}^{2}\{\mu(t)\} = {\bf D}^{2}\{\mu_{\ell}(t)\} +
{\bf D}^{2}\{\mu_{d}(t)\} + 2\;{\bf
Cov}\{\mu_{\ell}(t)\;\mu_{d}(t)\}, \] where
\[ {\bf Cov}\{\mu_{\ell}(t)\;\mu_{d}(t)\} = {\bf E}\{[\mu_{\ell}(t)-m_{1}^{(\ell)}(t)]\;
[\mu_{d}(t)-m_{1}^{(d)}(t)]\}, \] and so we have
\[ {\bf Cov}\{\mu_{\ell}(t)\;\mu_{d}(t)\} =
\frac{1}{2}\left[{\bf D}^{2}\{\mu(t)\} - {\bf
D}^{2}\{\mu_{\ell}(t)\} - {\bf D}^{2}\{\mu_{d}(t)\}\right]. \] By
using Eqs. (\ref{46}), (\ref{29}) and (\ref{38}) we find
\[ {\bf Cov}\{\mu_{\ell}(t)\;\mu_{d}(t)\} = \frac{{\bf D}^{2}\{\nu\}}{1-q_{1}}\;Qt\;
e^{-(1-q_{1})Qt} - \]
\begin{equation} \label{53}
\left[1 + \frac{{\bf
D}^{2}\{\nu\}}{(1-q_{1})^{2}}\right]\;e^{-(1-q_{1})Qt} \left[1 -
e^{-(1-q_{1})Qt}\right],
\end{equation}
\[ \mbox{if} \;\;\;\; q_{1} \neq 1. \]
In the case of critical evolution by taking into account the
expressions (\ref{47}), (\ref{30}), and (\ref{39}) the covariance
can be written in the form:
\begin{equation} \label{54}
{\bf Cov}\{\mu_{\ell}(t)\;\mu_{d}(t)\} =
\frac{1}{2}\;q_{2}\;(Qt)^{2}.
\end{equation}
These expression (\ref{53}) is exactly the same as Eq. (\ref{51})
while (\ref{54}) coincides with (\ref{52}).

\begin{figure} [ht]
\protect \centering{
\includegraphics[height=8cm, width=10cm]{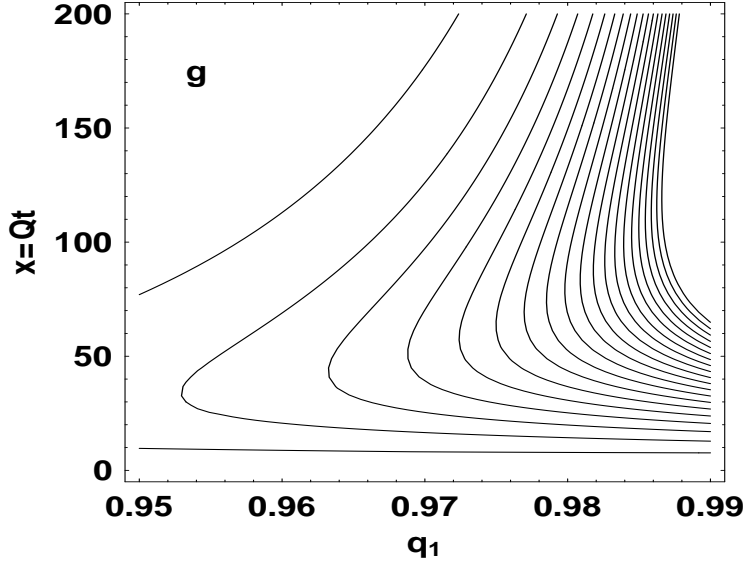}}\protect
\vskip 0.2cm \protect \caption{\footnotesize{Contour plot of the
covariance function ${\bf Cov}\{\mu_{\ell}(t) \mu_{d}(t)\}=
{\mathcal D}^{(\ell,d)}(Qt, q_{1})$ in the case of subcritical
random evolution by assuming that the offspring number $\nu$ is of
geometric distribution.}} \label{fig5}
\end{figure}

\begin{figure} [ht]
\protect \centering{
\includegraphics[height=8cm, width=12cm]{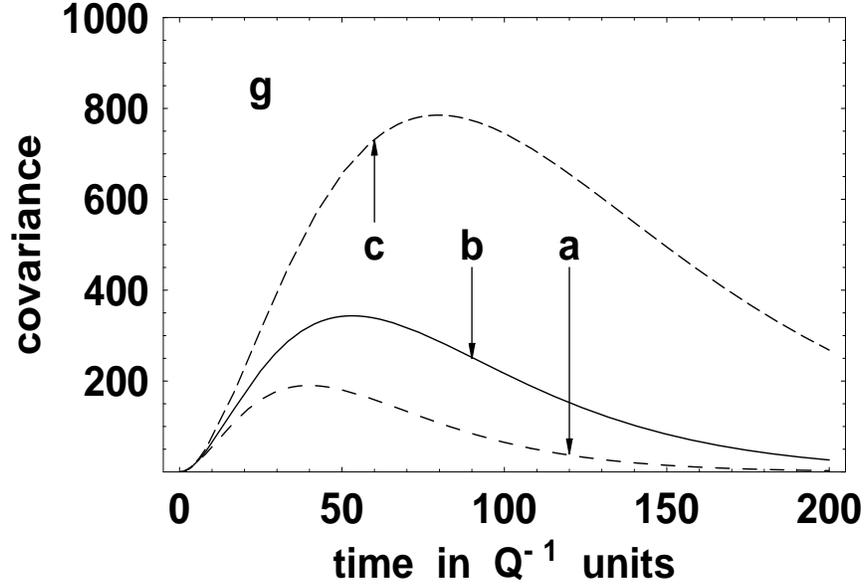}}\protect
\vskip 0.2cm \protect \caption{\footnotesize{Three curves
illustrating the time dependence of the covariance ${\mathcal
D}^{(\ell,d)}(Qt, q_{1})$ in the case of subcritical random
evolution by assuming that the $\nu$ is of geometric distribution.
The curves {\bf a, b}, and {\bf c} correspond to the values of
$q_{1}=0.96, 0.97$ and $0.98$, respectively.}} \label{fig6}
\end{figure}
\begin{figure} [!ht]
\protect \centering{
\includegraphics[height=5.5cm, width=8cm]{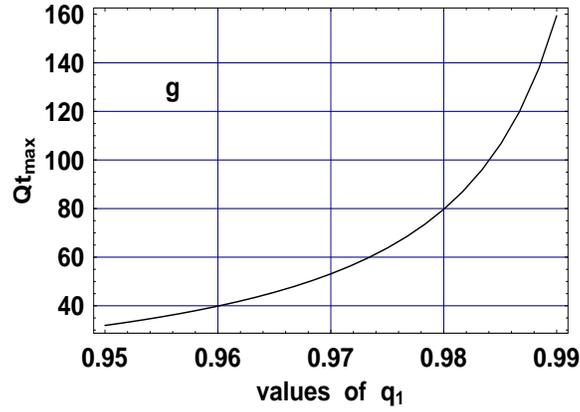}}\protect
\vskip 0.2cm \protect \caption{\footnotesize{The dependence of
$Qt_{m}$ on $q_{1}$ in the case of subcritical evolution assuming
that the $\nu$ is of geometric distribution.}} \label{fig7}
\end{figure}

\begin{figure} [ht]
\protect \centering{
\includegraphics[height=8cm, width=12cm]{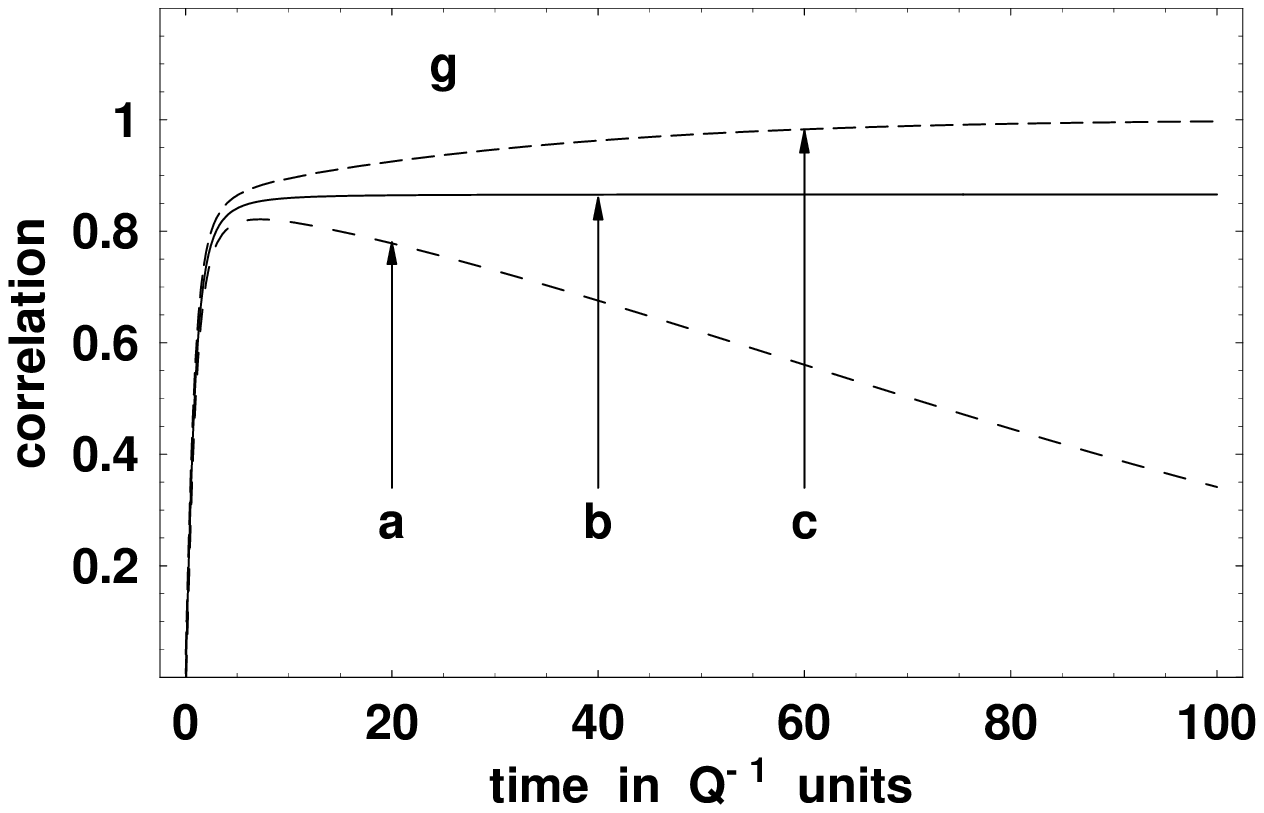}}\protect
\vskip 0.2cm \protect \caption{\footnotesize{Time dependence of
the correlation between the numbers living and dead nodes if $\nu$
is of geometric distribution. The curves {\bf a, b,} and {\bf c}
belong to $q_{1}$ values $0.95, \; 1,$ and $1.05$, respectively.}}
\label{fig8}
\end{figure}

The dependence of the covariance function \[{\bf
Cov}\{\mu_{\ell}(t) \mu_{d}(t)\}= {\mathcal D}^{(\ell,d)}(Qt,\;
{\bf E}\{\nu\},\; {\bf D}^{2}\{\nu\})\] on the time parameter $Qt$
and the expectation value as well as the variance of the $\nu$ has
been calculated by {\em assuming the $\nu$ to be of geometric
distribution}. Since in this case ${\bf E}\{\nu\} = q_{1}$ and
${\bf D}^{2}\{\nu\} = q_{1}(1+q_{1})$, thus one can write
\[ {\bf Cov}\{\mu_{\ell}(t) \mu_{d}(t)\}= {\mathcal D}^{(\ell,d)}(Qt,
q_{1}). \] The contour plot of this covariance function has been
calculated in the $(Qt, q_{1})$ plain for the subcritical values
of $q_{1}$. The results are seen in Fig. \ref{5}. In order to show
more precisely the form of time dependence of ${\mathcal
D}^{(\ell,d)}(x, q_{1})$, three curves {\bf a, b}, and {\bf c}
corresponding to the values $q_{1}=0.96, 0.97$, and $0.98$,
respectively, are plotted in Fig. \ref{6}. One can observe that
each of the covariance curves vs. time has a well defined maximum.
The time parameter $Qt_{m}$ due to the maximum of the covariance
function depends rather sensitively on $q_{1}$.  It is interesting
to look at the variation of $Qt_{m}$ when $q_{1} \Rightarrow 1$.
The curve plotted in Fig. \ref{fig7} shows this variation, and one
can see a steep increase of $Qt_{m}$ when $q_{1}$ is approaching
to $1$.

\begin{figure} [ht]
\protect \centering{
\includegraphics[height=8cm, width=12cm]{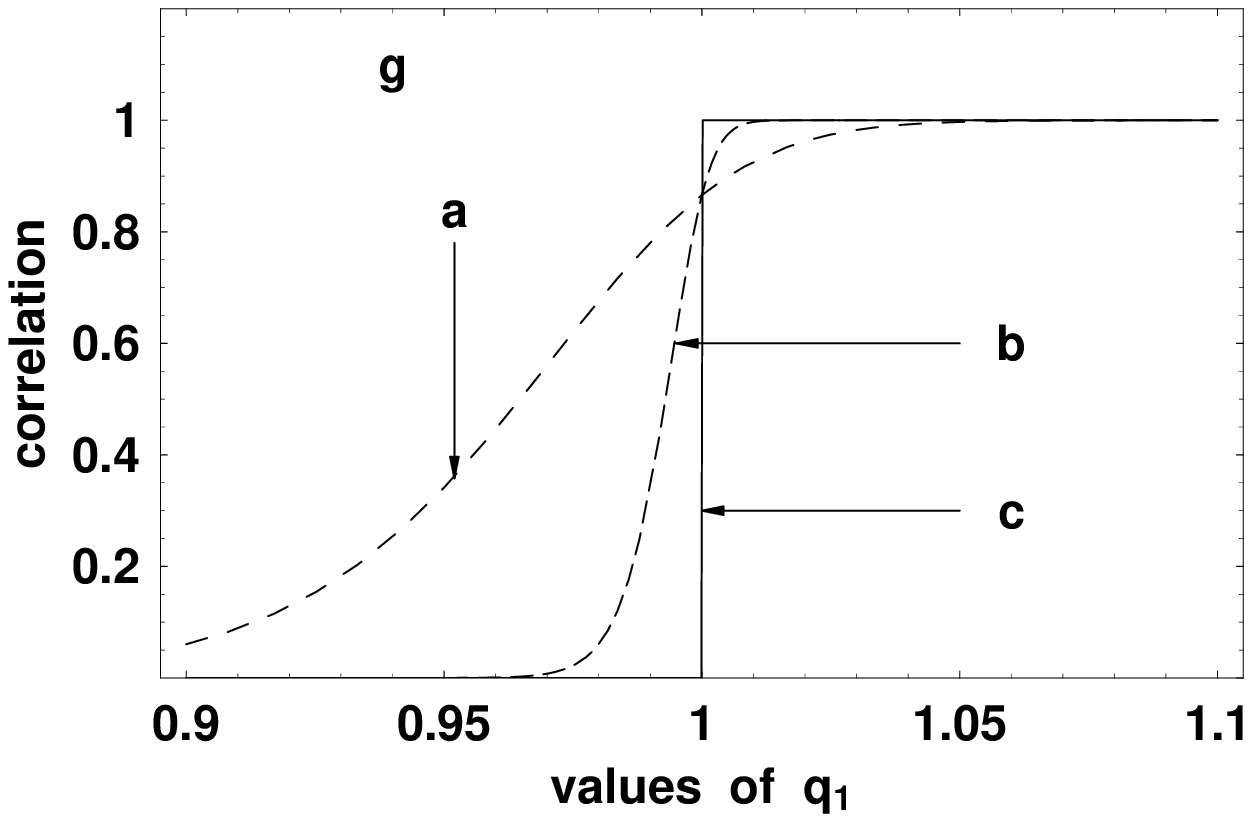}}\protect
\vskip 0.2cm \protect \caption{\footnotesize{Dependence of the
correlation between the numbers of living and dead nodes on
$q_{1}$ at time instants $Qt=100$, ({\bf a}), $Qtx=500$, ({\bf b})
and $Qt=\infty$, ({\bf c}).}} \label{fig9}
\end{figure}

{\bf Correlation function.} \hspace{0.2cm} It is trivial that the
numbers of living and dead nodes are correlated, but the time
dependence of the correlation function
\begin{equation} \label{55}
{\mathcal C}^{(\ell,d)}(t, q_{1}, q_{2}) = \frac{{\bf
Cov}\{\mu_{\ell}(t) \mu_{d}(t)\}}{{\bf D}\{\mu_{\ell}(t)\} {\bf
D}\{\mu_{d}(t)\}},
\end{equation}
has some specific features, namely, in the case of subcritical
evolution the correlation function --- after a sharp increase ---
is converging to zero when $Qt \Rightarrow \infty$, while in the
case of supercritical evolution reaching  rapidly the value $1$.

If the random evolution is critical, i.e. $q_{1}=1$, then we have
\begin{equation} \label{56}
{\mathcal C}^{(\ell,d)}(t, 1, q_{2}) = \frac{\sqrt{3}}{2}\;\left[1
+ \frac{3}{{\bf D}^{2}\{\nu\}\;(Qt)^{2}}\right]^{-1/2},
\end{equation}
which shows that the {\em limit value of the correlation}
\begin{equation} \label{57}
\lim_{Qt \rightarrow \infty}\;{\mathcal C}^{(\ell,d)}(t, 1, q_{2})
= \frac{\sqrt{3}}{2}
\end{equation}
is independent of distributions of $\nu$ and $\tau$. This very
important {\em limit law} expresses the fact that the relation
between the numbers of living and dead nodes near the critical
state, and at time moments sufficiently far from the beginning of
the process, becomes almost independent of rules controlling the
evolution.

Fig. \ref{fig8} \hspace{0.1cm} illustrates the time dependence
\hspace{0.1cm} of the correlation function $\;\;{\mathcal
C}^{(\ell,d)}(t, q_{1}, q_{2})$ in the case if $\nu$ is of
geometric distribution. The curves {\bf a, b,} and {\bf c} belong
to the $q_{1}$ values $0.95, \; 1,$ and $1.05$, respectively. It
can be shown that these curves are rather insensitive to the
distribution of $\nu$.

It seems to be worthwhile to present some curves reflecting the
dependence of ${\mathcal C}^{(\ell,d)}(t, q_{1}, q_{2})$ on
$q_{1}$ at different time instants. In Fig. \ref{9} we can see
that the transition of the correlation function from $0$ to $1$
becomes sharper and sharper with increasing $Qt$. The curves {\bf
a} and {\bf b} are belonging to the time parameters $Qt=100$ and
$Qt=500$, respectively, while the curve {\bf c} shows how  the
correlation function depends on $q_{1}$ at $Qt = \infty$. One has
to note that
\[ \lim_{Qt \rightarrow \infty}{\mathcal C}^{(\ell,d)}(t, 1+\epsilon,
q_{2}) = \left\{ \begin{array}{ll} 0, & \mbox{if $\epsilon < 0$,}
\\ \mbox{ } & \mbox{ } \\
\sqrt{3}/2, & \mbox{if $\epsilon = 0$,} \\
\\ \mbox{ } & \mbox{ } \\
1, & \mbox{if $\epsilon > 0$.}
\end{array} \right. \]
\subsubsection{Living nodes and lines}

In order to study the time dependence of correlation between the
numbers of living nodes and lines in randomly evolving trees we
should calculate the expectation value
\[ {\bf E}\{\mu_{\ell}(t)\;\mu_{e}(t)\} = m_{1,1}^{(\ell,e)}(t),
\] which can be determined from the generating function
$g^{(\ell,e)}(t,z_{\ell}, z_{e})$ defined by (\ref{21}). Since
\[m_{1,1}^{(\ell,e)}(t) = \left[\frac{\partial^{2}g^{(\ell,e)}(t,z_{\ell},
z_{e})}{\partial z_{\ell}\;\partial
z_{e}}\right]_{z_{\ell}=z_{e}=1},
\] one obtains from Eq. (\ref{22}) the integral equation
\[ m_{1,1}^{(\ell,e)}(t) =
Q\;q_{1}\;\int_{0}^{t}e^{-Q(t-t')}\;m_{1,1}^{(\ell,e)}(t')\;dt' +
\]
\begin{equation} \label{58}
Q\;\int_{0}^{t}e^{-Q(t-t')}\;m_{1}^{(\ell)}(t')\left\{q_{1} +
q_{2}\left[1 + m_{1}^{(e)}(t')\right]\right\}\;dt'.
\end{equation}
Taking into account Eqs.
\[m_{1}^{(\ell)}(t) = e^{-(1-q_{1})Qt} \]
and
\[ m_{1}^{(e)}(t) = \left\{ \begin{array}{ll}
        \frac{q_{1}}{1-q_{1}}\;[1-e^{-(1-q_{1})Qt}], & \mbox{if
        $q_{1} \neq 1$,} \\
        \mbox{ } & \mbox{ } \\
        Qt, & \mbox{if $q_{1}=1$,} \end{array} \right. \]
it can be shown that the solution of (\ref{58}) has the following
form:
\begin{equation} \label{59}
m_{1,1}^{(\ell,e)}(t) = \frac{{\bf
D}^{2}\{\nu\}}{1-q_{1}}\;Qt\;e^{-\alpha t} -
q_{2}\;\frac{q_{1}}{(1-q_{1})^{2}}\;e^{-\alpha t}(1 - e^{-\alpha
t}),
\end{equation}
\[  \mbox{if} \;\;\;\;\;\; q_{1} \neq 1, \]
where $\alpha = (1-q_{1})Q$, and
\begin{equation} \label{60}
m_{1,1}^{(\ell,e)}(t) = (1+q_{2})\;Qt +
\frac{1}{2}\;q_{2}\;(Qt)^{2}, \;\;\;\;\;\; \mbox{if} \;\;\;\;\;\;
q_{1} = 1.
\end{equation}
By using the relation
\[ {\bf Cov}\{\mu_{\ell}(t)\;\mu_{e}(t)\} = m_{1,1}^{(\ell,e)}(t)
- m_{1}^{(\ell)}(t)\;m_{1}^{(e)}(t) \] one can obtain that
\begin{equation} \label{61}
{\bf Cov}\{\mu_{\ell}(t)\;\mu_{e}(t)\} = \frac{{\bf
D}^{2}\{\nu\}}{1-q_{1}}\;Qt\;e^{-\alpha t} - q_{1}\left[1 +
\frac{{\bf D}^{2}\{\nu\}}{(1-q_{1})^{2}}\right]\;e^{-\alpha t}(1 -
e^{-\alpha t}),
\end{equation}
\[ \mbox{if} \;\;\;\;\;\; q_{1} \neq 1, \]
and
\begin{equation} \label{62}
{\bf Cov}\{\mu_{\ell}(t)\;\mu_{e}(t)\} = {\bf
D}^{2}\{\nu\}\;Qt\left(1 + \frac{1}{2}\;Qt\right),
\end{equation}
\[  \mbox{if} \;\;\;\;\;\; q_{1}=1. \]

\begin{figure} [ht]
\protect \centering{\includegraphics[height=8cm,
width=12cm]{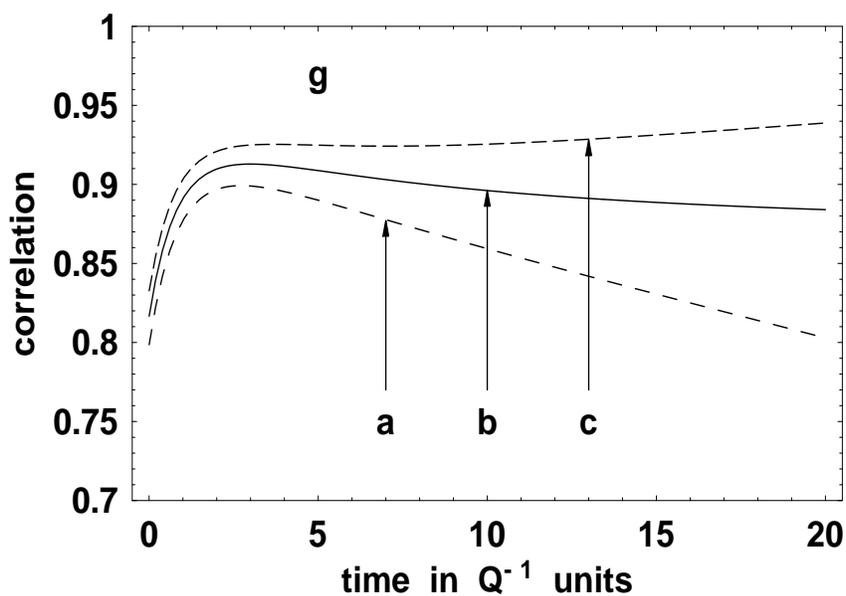}}\protect \vskip 0.2cm \protect
\caption{\footnotesize{Time dependence of the correlation between
the numbers of lines and living nodes in the case when $\nu$ is of
geometric distribution. The curves {\bf a, b,} and {\bf c} are
belonging to the $q_{1}$ values $0.95, \; 1,$ and $1.05$,
respectively.}} \label{fig10}
\end{figure}

Let us investigate the time dependence of the correlation function
\begin{equation} \label{63}
{\mathcal C}^{(\ell,e)}(t, q_{1}, q_{2}) = \frac{{\bf
Cov}\{\mu_{\ell}(t) \mu_{e}(t)\}}{{\bf D}\{\mu_{\ell}(t)\} {\bf
D}\{\mu_{e}(t)\}}.
\end{equation}
It can be proved that
\begin{equation} \label{64}
{\bf D}\{\mu_{e}(t)\} = {\bf D}\{\mu(t)\},
\end{equation}
and thus, if $q_{1} \neq 1$ then there is an easy task to
calculate (\ref{63}) by using the expressions (\ref{61}),
(\ref{29}) and (\ref{64}), while if $q_{1}=1$ then we can write
\begin{equation} \label{65}
{\mathcal C}^{(\ell,e)}(t, 1, q_{2}) = \frac{\sqrt{3}}{2}\;\frac{1
+ 2\;(Qt)^{-1}}{\left[1 + 3\;(Qt)^{-1}\left(1+\frac{1+{\bf
D^{2}}\{\nu\}}{{\bf
D}^{2}\{\nu\}}\;(Qt)^{-1}\right)\right]^{1/2}},
\end{equation}
from that it follows the limit relation
\[ \lim_{Qt \rightarrow \infty}{\mathcal
C}^{(\ell,e)}(t, 1, q_{2}) = \frac{\sqrt{3}}{2}, \] which is the
same as $\lim_{t \rightarrow \infty}{\mathcal C}^{(\ell,d)}(t, 1,
q_{2})$ and so, we can conclude that the limit value of the
correlation between the numbers of lines and living nodes is
completely independent of the parameters determining the tree
evolution process.

Fig. \ref{fig10} shows the time dependence of the correlation
${\mathcal C}^{(\ell,e)}(t, q_{1}, q_{2})$ near the beginning of
the process at the values of $q_{1}=0.95, 1, 1.05$ assuming that
$\nu$ has geometric distribution. The curve {\bf a} corresponding
to the subcritical evolution has a well defined maximum, and after
that it decreases rapidly to zero when $t \Rightarrow \infty$. In
the case of the critical evolution, i.e., when $q_{1}=1$, the
correlation function passing its maximum converges to the limit
value $\sqrt{3}/2$ as seen on the curve {\bf b}. Finally, one can
see that the curve {\bf c} due to the value $q_{1}=1.05$ is
reaching the level $1$ not monotonously but through a slight
maximum. The dependence of the correlation function ${\mathcal
C}^{(\ell,e)}(t, q_{1}, q_{2})$ on $q_{1}$ is similar to that of
${\mathcal C}^{(\ell,d)}(t, q_{1}, q_{2})$, i.e., the transition
from $0$ to $1$ becomes sharper and sharper with increasing $Qt$.

\section{Concluding remarks}

By introducing the notions of living and dead nodes a new model of
random tree evolution with continuous time parameter has been
constructed . In order to describe the evolution process two basic
random functions $\mu_{\ell}(t)$ and $\mu_{d}(t)$ have been used.
The first is the actual number of living nodes, while the second
that of the dead nodes at time instant $t \geq 0$.

Having assumed the evolution process to be controlled by the
lifetime and the offspring number of living nodes we derived exact
equations for generating functions of $\mu_{\ell}(t), \;\;
\mu_{d}(t),\;\; \{\mu_{\ell}(t),\mu_{d}(t)\}$ etc., provided that
at $t=0$ the tree consisted of a single living node only. It is
remarkable that the average lifetime has a role of scaling the
time only in these equations.

The time dependence of the average number of nodes has been used
for characterization of the evolution which can be either
subcritical or critical, or supercritical. It has been proved that
the relative variance of the tree size is increasing with time,
ie. at large $t$ the average tree size is hardly informative. If
the average offspring number converges to $1$, then it has been
found that the limit value of the relative variance tends to
infinity.

A specific property of the tree evolution has been discovered,
namely, it has been proved that the correlation between the
numbers of living and dead nodes decreases to zero in subcritical
and increases to $1$ in supercritical evolution, if the time
parameter $Qt$ tends to infinity, but in the case of exactly
critical evolution it converges to a fixed value $\sqrt{3}/2$
which is free of the parameters of the process.


\begin{thebibliography}{99}
\bibitem{barabasi01} R. Albert and A.-L. Barab\'asi, Statistical
Mechanics of Complex Networks, {\bf cond-mat/0106096}
\bibitem{dorogovtsev01} S.N. Dorogovtsev and J.F. Mendes,
Evolution of Random Networks, {\bf cond-mat/0106144}
\bibitem{erdos59} P. Erd\"os and A. R\'enyi, Publicationes Mathematicea {\bf 6} (1959) 290
\bibitem{erdos60} P. Erd\"os and A. R\'enyi, Math. Inst. Hung. Acsad. {\bf 5} (1960) 17
\bibitem{erdos61} P. Erd\"os and A. R\'enyi, Bull. Int. Stat. Inst, {\bf 38} (1961) 343
\bibitem{watts98} D.J. Watts and S.H. Strogatz, Nature {\bf 393} (1998) 440
\bibitem{watts99} D.J. Watts, {\em The Dynamics of Networks Between Order and Randomness}
(Princeton University Press, Princeton New-Jersey, 1999)
\bibitem{newman00} M.E. Newman, C. Moore, and D.J. Watts, Phys. Rev. Lett.
{\bf 84} (2000) 3201
\bibitem{barabasi99} A. Barab\'asi, R. Albert, and H. Jeong, Phisica A
{\bf 272} (1999) 173
\bibitem{krapivsky00} P.L. Krapivsky, S. Redner, and L. Leyvraz, Phys. Rev. Lett.
{\bf 85} (2000) 4629
\bibitem{dorogovtsev00} S.N. Dorogovtsev, J.F. Mendes and A. Samukhin,
{\bf cond-mat/0004434}
\bibitem{harris63} T.E. Harris, {\em The theory of Branching Processes}
(Springer-Verlag, Berlin-G\"ottingen-Heidelberg, 1963)
\bibitem{lyons02} R. Lyons and Y. Peres, {\em Probability on Trees
and Networks}, (Cambridge University Press, in preparation,
Current version available at {\bf http:/www. math.
washington.edu/~ejpecp/})
\bibitem{sevast'yanov71} B. Sewastjanow, {\em Verzweigungsprozesse} (Akademie-Verlag, Berlin, 1974)
\bibitem{lpal95} L. P\'al, {\em The Foundation of the Probability
Calculus and Statistics} (Akad\'e\-miai Kiad\'o, Budapest, 1995)
\bibitem{lpal94} L. P\'al, {\em Properties of Generating Functions}
(Internal Report, Budapest, 1994)
\end{thebibliography}
\end{document}